\documentclass[journal,twoside]{IEEEtran}
\usepackage[sort,compress]{cite}

\RequirePackage{xspace}
\def\kaos{{\sc Kaos}\xspace}

\usepackage{graphicx}
\graphicspath{{figures/}}

\usepackage[caption=false,font=footnotesize]{subfig}

\begin{document}
\title{A Large-Scale FPGA-Based Trigger\\ and Dead-Time Free DAQ
  System\\ for the {\sc Kaos} Spectrometer at MAMI}

\author{P.~Achenbach,
	C.~Ayerbe Gayoso, 
	J.~C.~Bernauer, 
	R.~B\"ohm,
	D.~Bosnar,
	L.~Debenjak, 
  M.~O.~Distler, 
  A.~Esser,
  I.~Fri{\v s}{\v c}i{\'c},
  M.~{G\'omez~Rodr\'iguez~de~la~Paz},
  J.~Hoffmann,
  M.~Makek,
  H.~Merkel, 
  S.~Minami,
  U.~M\"uller, 
  L.~Nungesser,
  W.~Ott,
  J.~Pochodzalla, 
  M.~Potokar,
  I.~Rusanov,
	T.~R.~Saito,
  S.~{S\'anchez Majos}, 
  B.~S.~Schlimme,
	S.~\v Sirca,
  S.~Voltz,
  K.~Weindel,
  M.~Weinriefer%
\thanks{Manuscript received June 4, 2010; revised May 1, 2011. This work was supported in
  part by Bundesministerium f{\"u}r Bildung und Forschung (bmb+f)
  under Contract 06MZ176, by the Federal State of Rhineland-Palatinate
  and the Deutsche Forschungsgemeinschaft with the Collaborative
  Research Center 443, and by the GSI as F\&E project MZ/POC.}%
\thanks{P.~Achenbach,
	C.~Ayerbe Gayoso, 
	J.~C.~Bernauer, 
	R.~B\"ohm,
  M.~O.~Distler, 
  A.~Esser,
  M.~{G\'omez~Rodr\'iguez~de~la~Paz},
  H.~Merkel, 
  U.~M\"uller, 
  L.~Nungesser,
  J.~Pochodzalla,
  S.~{S\'anchez Majos}, 
  B.~S.~Schlimme, 
  K.~Weindel, and
  M.~Weinriefer
are with the Institut f\"ur Kernphysik, Johannes
Gutenberg-Universit\"at, Mainz, Germany (telephone: +49-6131-39-25831,
e-mail: patrick@kph.uni-mainz.de).}%
\thanks{
  J.~Hoffmann,
  S. Minami,
  W.~Ott,
  I.~Rusanov,
  T. R. Saito,
  and S.~Voltz are with the GSI Helmholtzzentrum
  f\"ur Schwerionenforschung, Darmstadt, Germany.}%
\thanks{D.~Bosnar, I. Fri{\v s}{\v c}i{\'c}, and M.~Makek are with the
  Department of Physics, University of Zagreb, Croatia.}%
\thanks{L.~Debenjak, M.~Potokar, and S.~\v Sirca are with the
  University of Ljubljana and Jo\v zef Stefan Institute, Ljubljana,
  Slovenia.}%
\thanks{Color versions of one or more of the figures in this paper are
  available online at http://ieeexplore.ieee.org.}
}

\maketitle

\begin{abstract}
  The \kaos\ spectrometer is maintained by the A1 collaboration at the
  Mainz Microtron MAMI with a focus on the study of $(e,e'K^+)$
  coincidence reactions. For its electron-arm two vertical planes of
  fiber arrays, each comprising approximately 10\,000 fibers, are
  operated close to zero degree scattering angle and in close
  proximity to the electron beam.  A nearly dead-time free DAQ system
  to acquire timing and tracking information has been installed for
  this spectrometer arm. The signals of 144 multi-anode
  photomultipliers are collected by 96-channel front-end boards,
  digitized by double-threshold discriminators and the signal time is
  picked up by state-of-the-art {\cal F}1 time-to-digital converter
  chips.  In order to minimize background rates a sophisticated
  trigger logic was implemented in newly developed {\sc Vuprom}
  modules. The trigger performs noise suppression, signal cluster
  finding, particle tracking, and coincidence timing, and can be
  expanded for kinematical matching $(e'K^+)$ coincidences. The full
  system was designed to process more than 4\,000 read-out channels
  and to cope with the high electron flux in the spectrometer and the
  high count rate requirement of the detectors. It was successfully
  in-beam tested at MAMI in 2009.
\end{abstract}

\section{Introduction}
\IEEEPARstart{S}{ince} 2008 the magnetic spectrometer \kaos, dedicated
to the detection of charged kaons, has operated at the 1.5\,GeV
electron beam of the Mainz Microtron MAMI at the Institut f\"ur
Kernphysik in Mainz, Germany.  The spectrometer is maintained by the
A1 collaboration with a focus on the study of $(e,e'K^+)$ coincidence
reactions.

A new coordinate detector has been developed for the
\kaos\ spectrometer's electron-arm.  The detector consists of two
vertical planes (called $x$ and $\theta$) of fiber arrays. The package
will be supplemented by one or two horizontal planes (called $y$ and
$\phi$). The track information is used to reconstruct the target
coordinates and particle momentum, and the time information is used to
determine the time-of-flight of the detected particle from target to
the detection planes.

The vertical detector system, comprising a total of 18\,432 fibers and
covering an active area of $L \times H \sim$ 1\,940\,mm $\times$
300\,mm, is divided into detector segments of 384 fibers each.  These
consist of four-layered arrays where columns of four fibers are
coupled to one common read-out channel of a multi-anode
photomultiplier (MaPMT). In the electron-arm of the spectrometer the
accepted particles are crossing the focal plane with an inclination
angle of 50--70$^\circ$ with respect to the normal of the plane.  To
accommodate this geometry the fiber arrays are constructed in a
hexagonal packing with columns slanted by 60$^\circ$.

\begin{figure}
  \centering
  \includegraphics[width=\columnwidth]{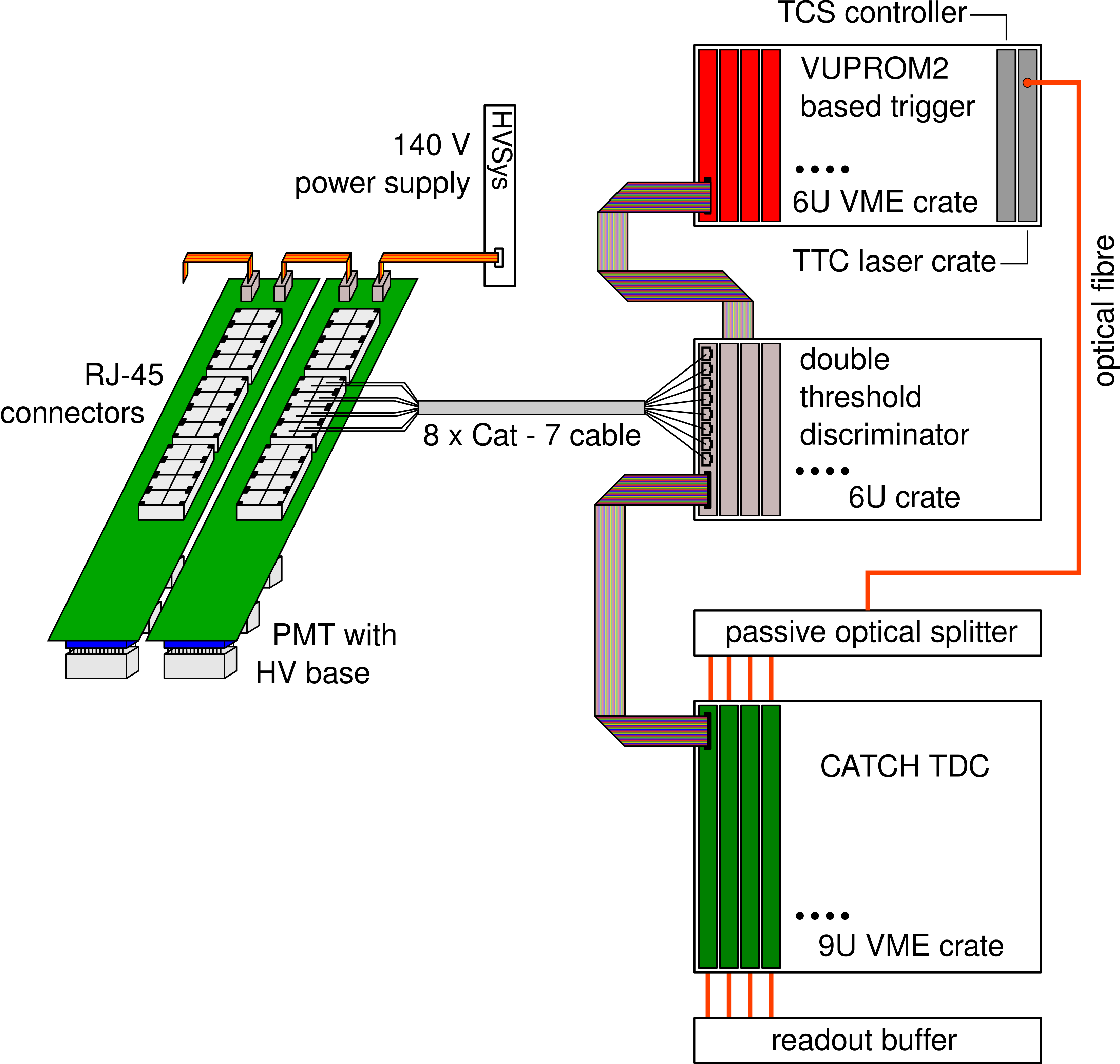}
  \caption{Scheme of the front-end, trigger, and data acquisition
    electronics for the electron-arm of the \kaos\ spectrometer at
    MAMI. The signals from three MaPMTs are collected with front-end
    boards and transferred over 15\,m to the discriminator modules
    (gray) which serve the {\sc Vuprom}2 trigger system (red) and the
    {\sc Catch} TDC system (green). The trigger is distributed to the
    {\sc Catch} modules by a laser system via optical fibers. The data
    is then transferred to the readout buffer PCs. The MaPMT high
    voltage is generated in the bases and is supplied to the
    daisy-chained front-end boards via a 140\,V line.}
  \label{fig:DAQ-scheme}
\end{figure}

For the double-arm operation of the spectrometer the 1.5\,GeV electron
beam must be steered through the \kaos spectrometer after impinging on
a nuclear target.  The fiber detectors are operated close to zero
degree scattering angle and in close proximity to the electron
beam. The electromagnetic background in the spectrometer is enormous
and for small detection angles is dominated by M{\o}ller
scattering. This background is considered to be the limiting factor
for the reaction rates that can be recorded.  Thus, the large number
of detector channels and the high particle fluxes require a flexible
and sophisticated trigger logic and a nearly dead-time free read-out
scheme.

\section{Fiber Detector Front-End}
For the fast signals of the 144 MaPMTs from the vertical planes of the
fiber detector in the electron arm new front-end electronics have been
developed. The bases of three neighboring MaPMTs are mounted directly
on 96-channel front-end boards.  These boards are used to provide the
appropriate voltages to the MaPMT bases and to distribute the analog
signals with a minimum of time jitter to eight RJ-45 connectors.  The
form-factor of the board follows the geometry of the fiber detector
that is aligned with the focal plane of the electron-arm of the
\kaos\ spectrometer. The signals (approx.\ 100\,mV pulse-height) are
then transported over 15\,m via Cat-7 patch cables that are
electromagnetically well shielded and that show small losses.  The
signal digitization is performed with double-threshold discrimination
(DTD) chips.  Eight chips, each digitizing 4 input channels, are
operated on a single 6U card that processes the signals of one MaPMT.
Details of the front-end cards and the DTD operation are described in
Ref.~\cite{Achenbach2009:FrontEndElectronics}. The cards have been
developed at the Electronics Department of the Institut f\"ur
Kernphysik at the Johannes Gutenberg-Universit\"at, Mainz. The output
signals are made available in LVDS standard on the front-side in a
68-pin Robinson-Nugent connector for the readout modules, and on the
back-side in a VHDCI connector for the trigger generation. The scheme
of the front-end, trigger, and data acquisition electronics is shown
in Fig.~\ref{fig:DAQ-scheme}.

\section{Trigger Logic}
A sophisticated trigger logic to be implemented in the FPGA has been
developed according to two different scenarios:
\begin{enumerate}
\item For the in-beam tests of the electron-arm detectors the trigger 
	system was based on a fast clustering
  algorithm to define the hit positions and thus the charged particle 
  track, and
  a hit multiplicity cut to reject spurious hits.  The trigger was
  prepared to minimize accidental trigger rates and to reject
  background events that do not originate in the target.
\item In hypernuclei electro-production experiments the trigger
  decision will also include the correlation between the momenta, {\em
    i.e.\/} the hit positions in the focal-planes of the
  spectrometers.  This sensitivity to the reaction kinematics allows
  to select a given missing-mass range by the trigger.  The temporal
  correlation between hits is used to identify and reject pile-up and
  to select specific particles via time-of-flight.  The possibility
  exists to expand the system to future additional detector components
  like the Cherenkov detector, which is under development.
\end{enumerate}

\begin{figure*}[!t]
  \centering
  \includegraphics[width=\columnwidth]{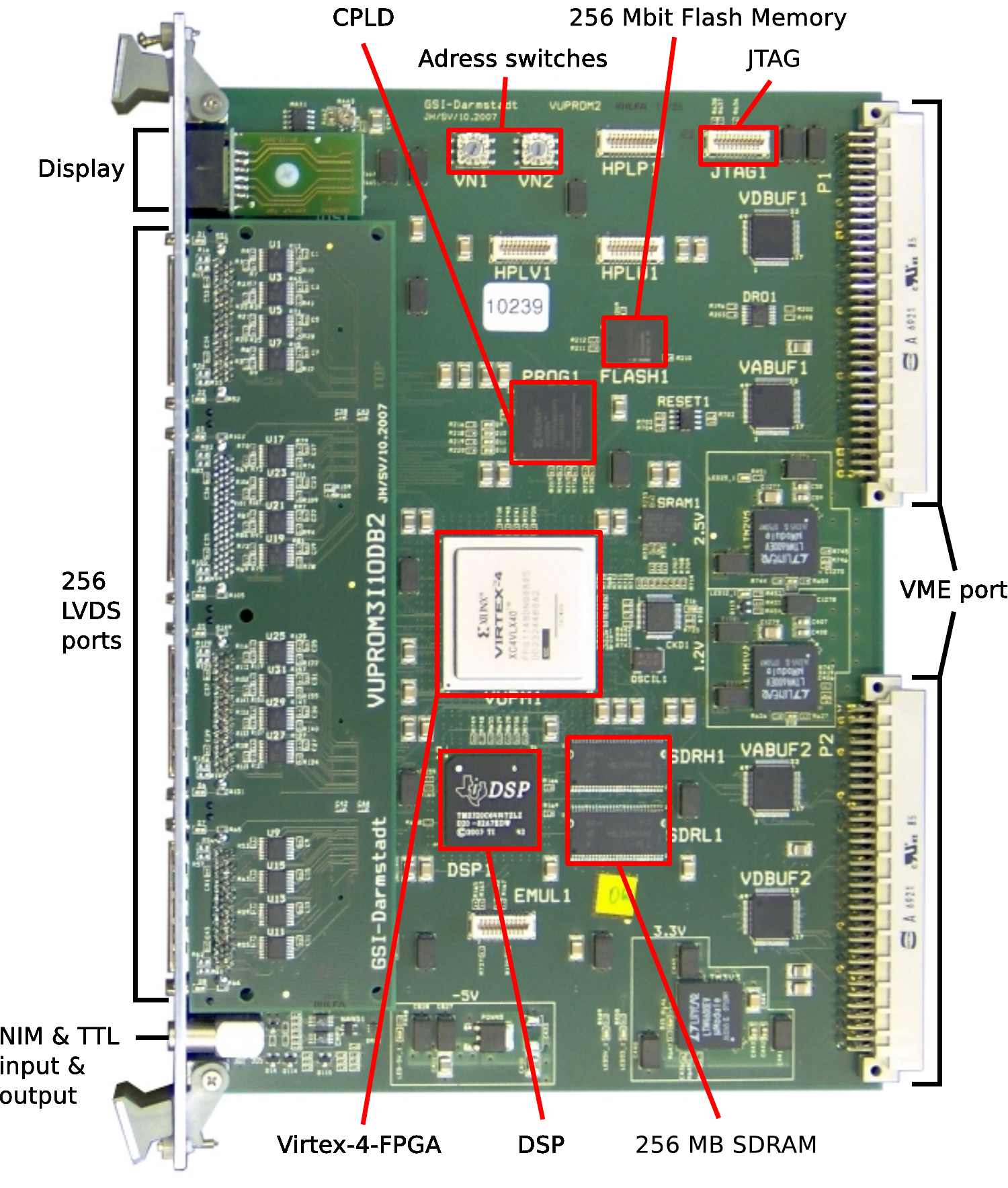}
  \caption{Photograph of a second generation logic module {\sc
      Vuprom}2 that is used in the trigger for the electron-arm of the
    \kaos\ spectrometer.  The following components are marked:
    {\sc Virtex-4} FPGA chip, CPLD chip, DSP chip, flash 
    memory, JTAG connector, piggy board, display, and  
    address switches.}
  \label{fig:VUPROM2}
\end{figure*}

Accordingly, the trigger must satisfy the following requirements:
\begin{enumerate}
\item detection of signal clusters in each detector plane;
\item reconstruction of particle tracks through both planes;
\item coincidence and acceptance test for all reconstructed particle tracks;
\item expandability to a missing-mass trigger by combination with the
  hadron-arm trigger;
\item on-line access and control of trigger parameters;
\item a fast trigger decision.
\end{enumerate}

\section{Vuprom Trigger System}
The trigger is derived by {\sc Vuprom} (Vme Universal PROcessing
Module) boards that were developed at the Experiment Electronics
Department of GSI for general purpose logic operations with 256 I$/$O
channels each~\cite{Hoffmann2008:GSI07,VUPROM2}.  A photograph of a
second generation {\sc Vuprom}2 module is shown in
Fig.~\ref{fig:VUPROM2}.  During the last 2 years, first modules were
applied as tracking triggers in experiments at GSI and at
MAMI~\cite{Minami2008:GSI07,Minami2009:GSI08,Sanchez2008:GSI07:VULOM}.

The first large-scale installation of a {\sc Vuprom} system with
several thousand input channels is now in operation for the triggering
of both the hadron arm and the electron arm instrumentation of the
\kaos\ spectrometer.  The electron-arm trigger setup comprises 37
{\sc Vuprom} modules.

Each VUPROM module has a 6U VME formfactor and is equipped with a Virtex-4 FPGA chip
containing over 40\,K logic cells, capable of operating at 400\,MHz
and connected to a DSP with 128\,Mbytes SDRAM.  The FPGA can be
accessed via a JTAG connector. For fast programming of all modules the
VME-bus is used to access on each board a 256\,Mbit flash memory that
is divided in 8 address spaces. After start-up or restart of the
module the FPGA configuration is read from two address spaces of the
flash memory, thus up to four independent configurations can be stored
on-board.  A CPLD supports the FPGA control and the flash memory
access for the FPGA configuration. The DSP is intended for complex
trigger calculations.  A display on the front panel is accessible from
the DSP and can be used to show status information of the module. The
trigger calculation results can be forwarded to front panel or to the
VME backplane bus.  Eight high density VHDCI connectors with 32
differential input or output channels each are implemented on the
front panel, with four connectors on the {\sc Vuprom} main board
hard-wired as three inputs and one output, and four piggyback options
with free configuration.  Additionally, standard LEMO in- and outputs
(TTL and NIM level) are available.  Thus, output and control
information is accessible during trigger operation and the trigger
system can be easily reconfigured and adopted to the experimental
requirements.

\section{Trigger Implementation}
\begin{figure*}[!t]
  \centering
  \includegraphics[width=0.7\textwidth]{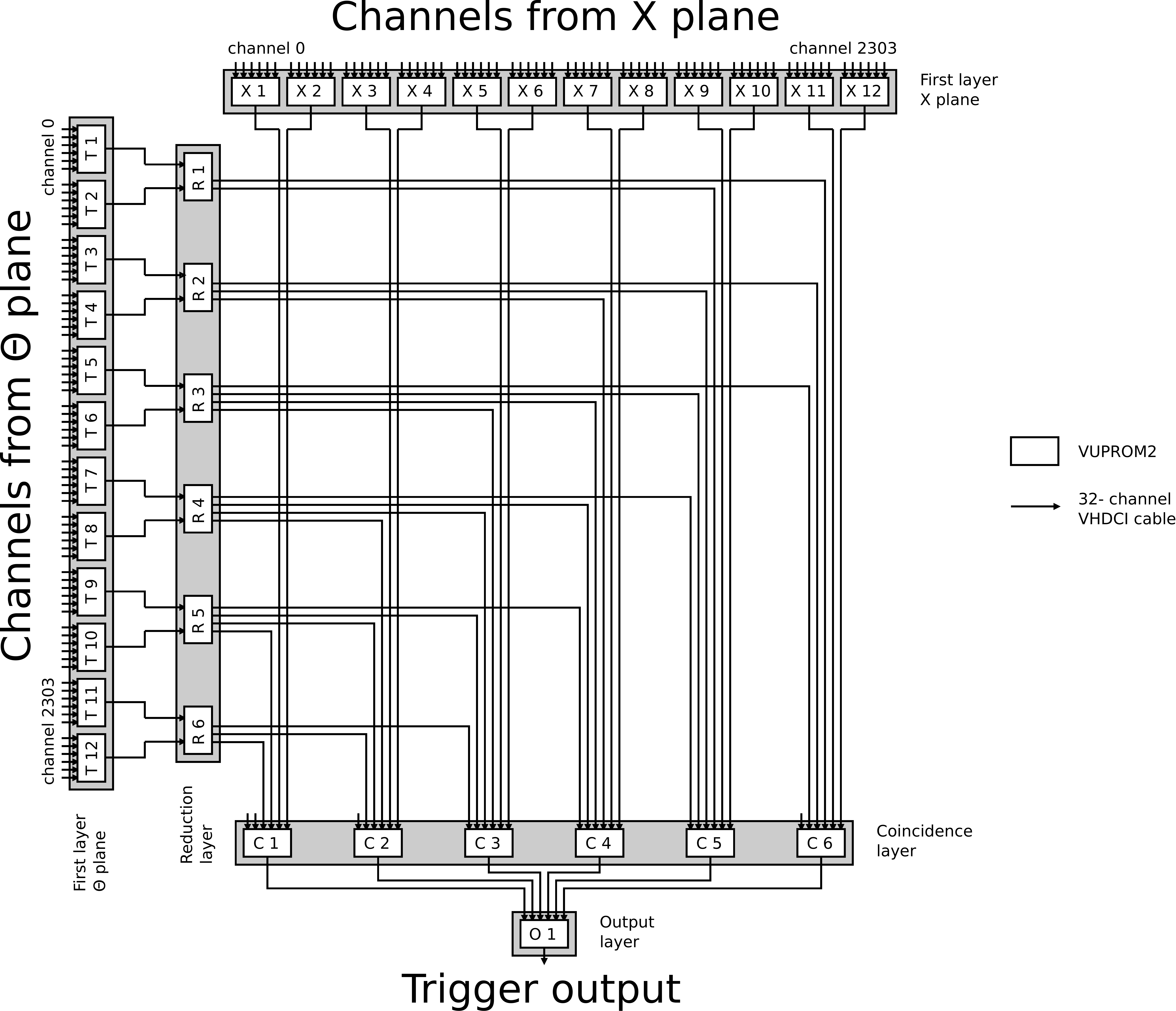}
  \caption{Trigger scheme used during the beam-test of the
    electron-arm detectors.  Each white box represents a {\sc Vuprom}2
    module, gray shaded areas represent the four different stages of
    the trigger logic, arrows indicate 32-channel VHDCI cables. Each
    fiber plane has its own first layer, the one for the
    $\theta$-plane is followed by a reduction layer, the coincidence
    and output layers combine the signals from the two planes. A total
    of 37 modules is needed to satisfy all the requirements on the
    trigger.}
  \label{fig:trigger-scheme}
\end{figure*}

\begin{figure*}[!t]
  \centering
  \subfloat[First stage]{
    \includegraphics[width=\columnwidth]{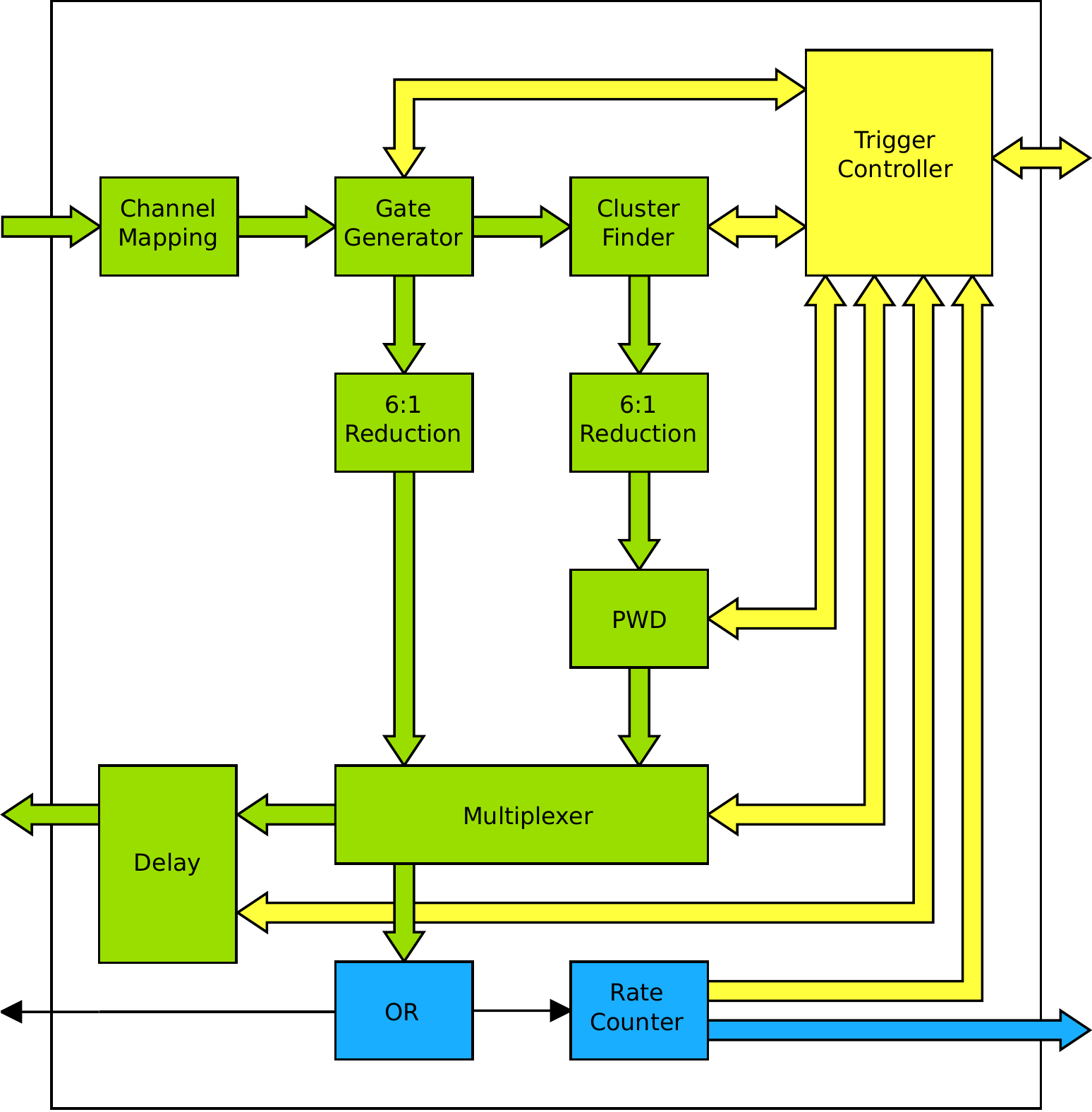}}\hfill
  \subfloat[Reduction stage]{
    \includegraphics[width=\columnwidth]{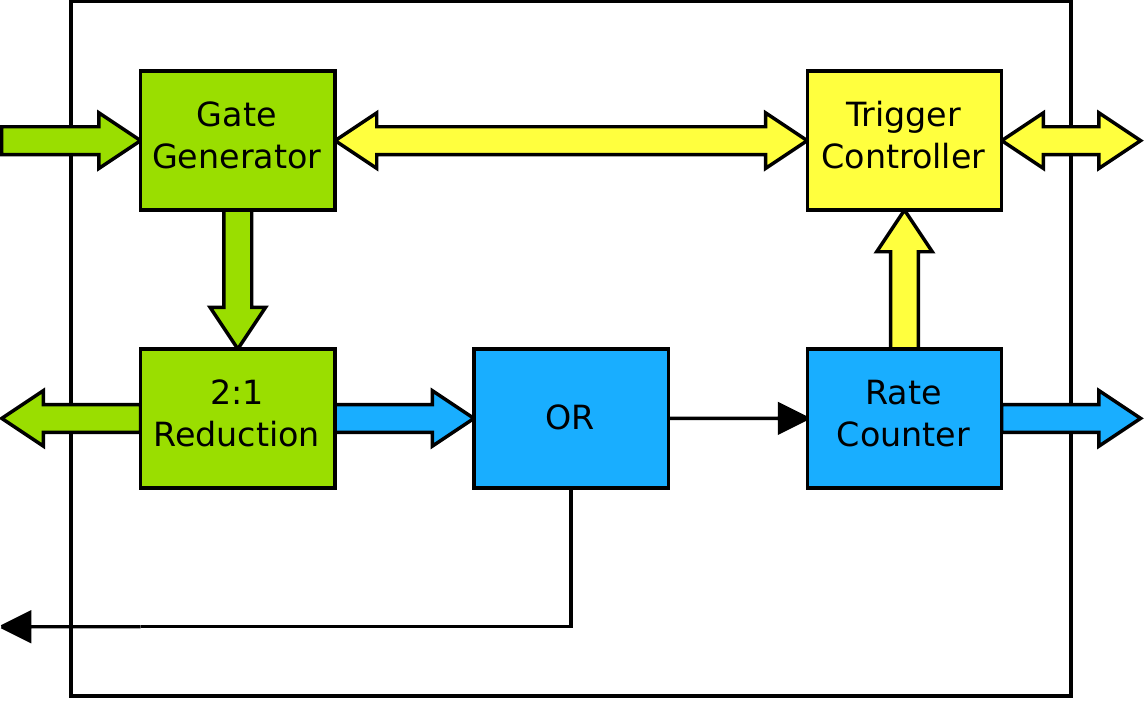}}\\
  \subfloat[Coincidence stage]{
    \includegraphics[width=\columnwidth]{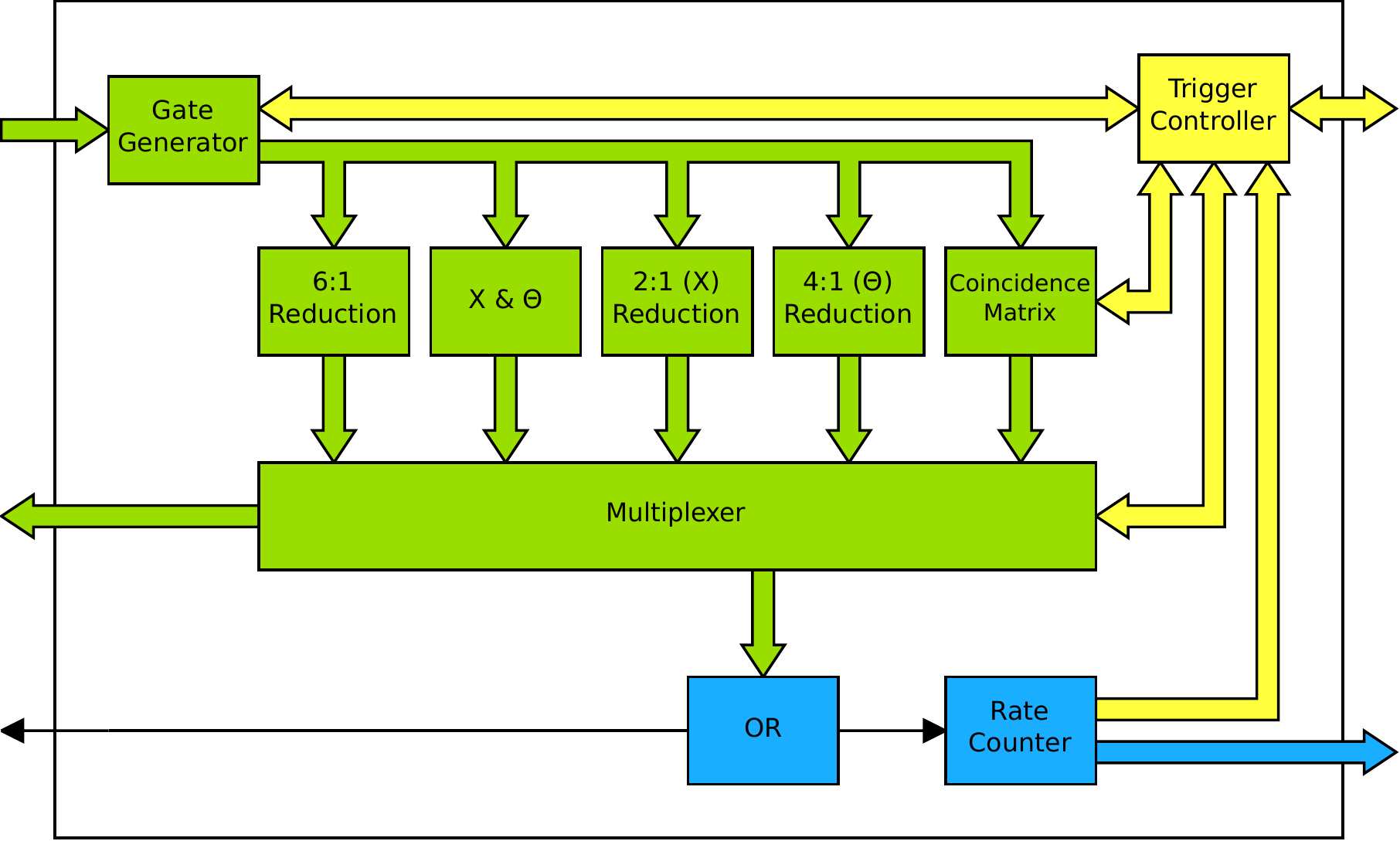}}\hfill
  \subfloat[Output stage]{
    \includegraphics[width=\columnwidth]{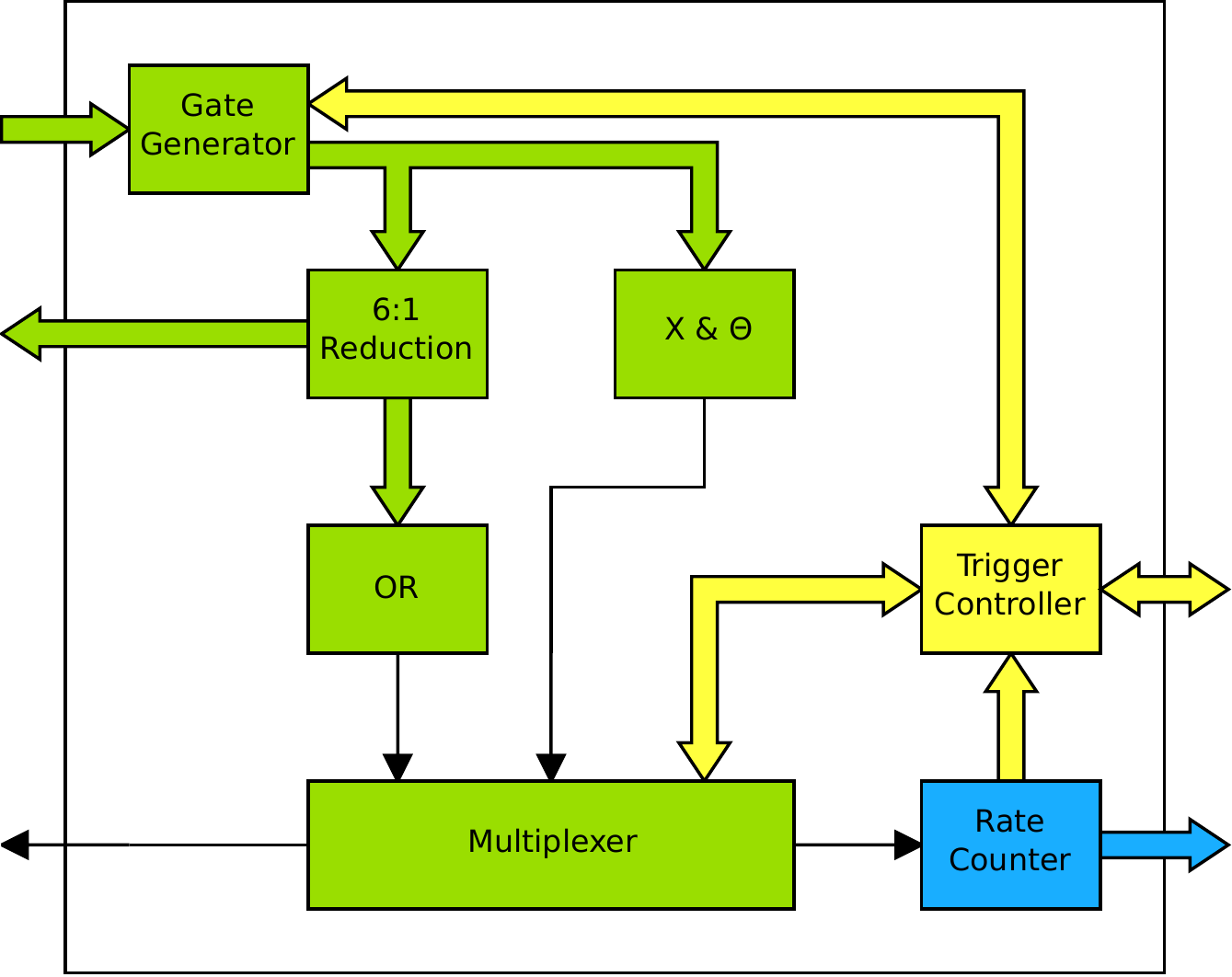}}
  \caption{Block diagrams of the implemented logic. Each box
    represents one functional unit, basic logic signals are represented
    as thin arrows, parallel signals as thick ones. The color code
    reflects the main purpose of the unit: green colored units and
    signals are part of trigger signal processing, yellow colored
    components are part of the trigger control, and debug or
    monitoring components are colored blue. (a) The first stage
    performs the channel mapping, the cluster finding, a channel
    reduction with pulse width discrimination, and a delayed output
    signal transmission. (b) The reduction stage performs a channel
    reduction by a factor of 2. (c) The coincidence stage consists of
    5 different logic units. (d) The output stage performs a temporal
    acceptance test, which can be combined with an angular acceptance
    test, and provides the first level trigger signal to the DAQ
    system. }
   \label{fig:logic}
\end{figure*}

A scheme of the trigger implementation is shown in Fig.~\ref{fig:trigger-scheme}.
The modules are arranged in four stages.  The signals from both
detector planes are processed in parallel and hit information is
transmitted to the next stage by setting bits in an 32-bit wide output
bus. The coincidence and output layers combine the signals from the
two planes. A total of 37 modules is needed to satisfy all the
requirements on the trigger. Block diagrams of the implemented logic
in all four stages are shown in Fig.~\ref{fig:logic}. Each box
represents one functional unit, thin arrows represent single logic
signals, parallel signals between components are represented as thick
arrows.  The color code reflects the main purpose of the component:
green colored units and signals are part of trigger signal processing,
yellow colored components are part of the trigger control, and debug
or monitoring components are colored blue.

\begin{figure}[!ht]
  \centering
  \includegraphics[width=0.5\columnwidth]{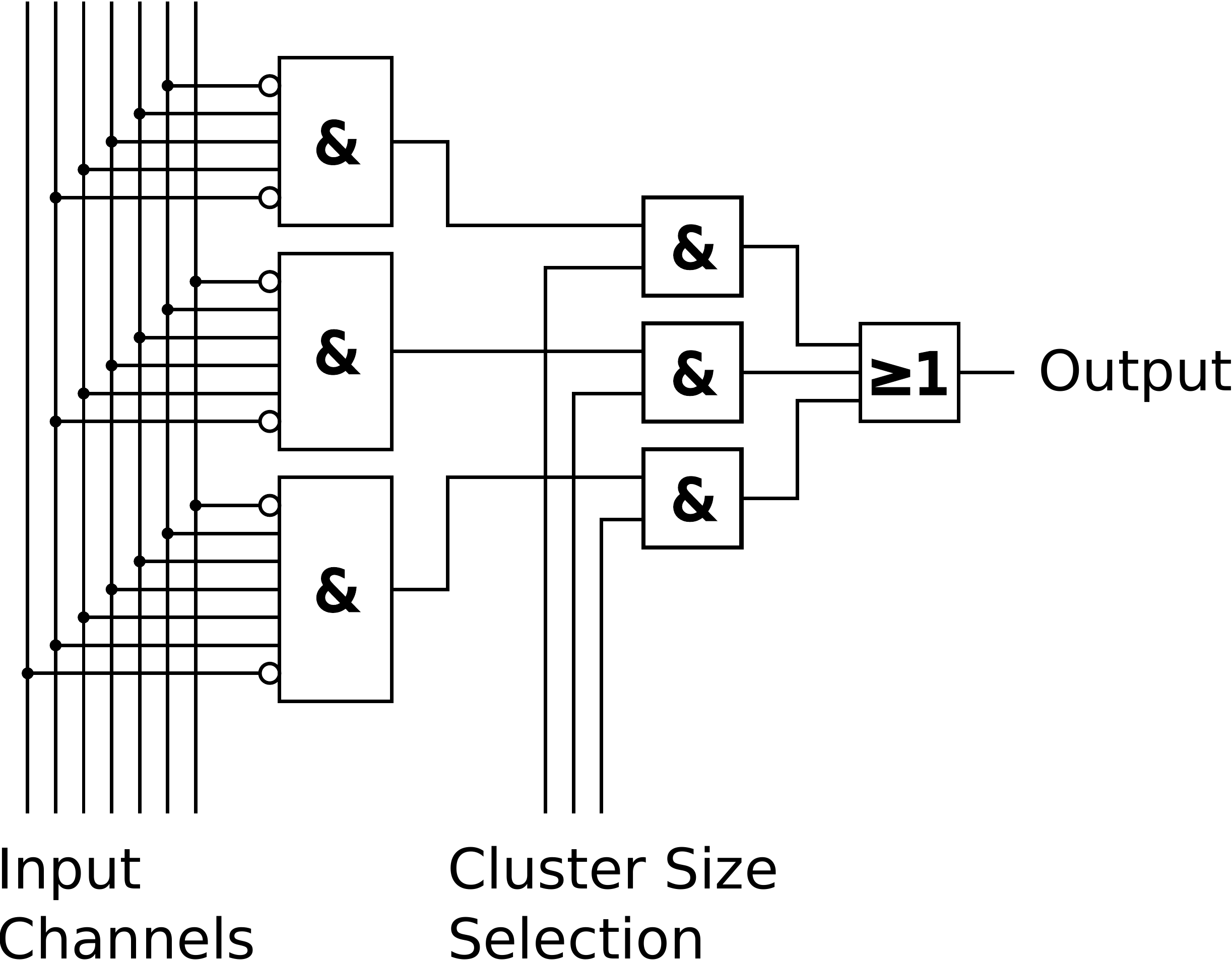}
  \caption{Detail of the logic unit for finding clusters of size $n=$
    3--5. The input signals are connected to
    AND gates, so that an output signal is only produced when a given
    signal size is found. Further information on the logic units is
    given in \cite{EsserDipl}.}
  \label{fig:clusterfinder}
\end{figure}

Each detector plane has 2\,304 trigger channels for which 72 VHDCI
inputs in the first stage are needed. This stage uses 12 {\sc Vuprom}
modules with 6 VHDCI inputs each for a single plane, with every module
serving 192 neighboring channels. One VHDCI output is used to transmit
information to the next stage, one VHDCI output is used for debugging.
Due to the geometrical arrangement of the fibers, a particle hit
always causes a cluster of correlated signals in neighboring
channels. Signal clusters are identified by requiring a signal in $n$
neighboring channels and the absence of a signal in the next higher
and lower channel within a given time period.  This scheme, shown in
Fig.~\ref{fig:clusterfinder} ensures that the found cluster is exactly
of the size $n$.  Short pulses which are the result of different
delays in the input signals are suppressed by a pulse-width
discriminator (PWD). The PWD splits the signal into two parts with one
part delayed. Both signals are then fed to an AND gate, such that only
signals with a length greater than the delay produce an output signal.
Clusters with a size above the given upper boundary are discarded to
reduce background from scattered particles, because these hit the
detector planes under large angles with respect to the normal on the
detector plane. Extremely small clusters can only be produced by
noise.  The upper and lower boundary for the accepted cluster sizes
can be set online via the VME bus and can be adapted to experimental
conditions.

The second stage consists of 6 modules with the purpose of reducing
the number of channels by a factor of 2 by executing an OR 
between neighboring channels and providing the signal
information on several outputs in parallel.

In the third stage the position information from the first stage of
the $x$-plane and the second stage of the $\theta$-plane are combined.
The stage consists of 5 different logic units and a multiplexer to
chose the logic for the output. A 6~:~1 reduction unit serves as an OR
gate for all channels. One unit checks for temporal coincidences
between the $x$- and $\theta$-planes.  A 2~:~1 reduction unit produces
an OR output for the channels coming from the $x$-plane. For
the $\theta$-plane a 4~:~1 reduction unit is implemented. As the outputs
of the reduction stage are connected to several inputs of the
coincidence stage, a signal of the $\theta$-plane can be processed in
different modules. 


If a correlation of the hit positions is required, a
coincidence matrix is used to perform an acceptance test for the
reconstructed tracks: The allowed combinations of hit
positions in both planes are stored in a binary matrix that was
externally computed.  An output signal is produced when a cluster
is found on both detector planes and the corresponding matrix element
is non-zero. Each {\sc Vuprom} in the third stage evaluates a range
of 384 channels of the $x$-plane and up to 1\,536 channels of the
$\theta$-plane.  This setup was chosen to cover the angular
acceptance known from simulations with the highest possible
granularity.  The position resolution for the acceptance test is 6
channels in the $x$-plane (4.98\,mm) and 12 channels in the
$\theta$-plane (9.96\,mm).

The output consists of a single module that receives the
information of accepted trajectories from the third stage and produces
the trigger output signal. Two parallel logic units are implemented,
one is a coincidence unit between signals of the $x$- and
$\theta$-planes, the other is an OR gate.
If a trigger is based on coincidences independent of the hit position, 
this decision needs to be taken by the output stage.

Because of the varying complexity of the four trigger stages, each 
one has a different latency. The input stage latency is 68\,ns, the
reduction stage latency 22\,ns, the coincidence stage latency 23--30\,ns 
depending on the selected trigger type, and the output stage latency is
30\,ns. The through-put time for raw signals is 104\,ns, the generation
of a $x$ AND $\theta$ trigger needs 143\,ns, and additional tracking
conditions require extra 7\,ns. 
There is no event-by-event variation of the latencies and 
a time jitter of $\pm$1.5\,ns is occurring in each stage, producing a 
total jitter of $\pm$3\,ns.

The fraction of occupied slices in the FPGA is 45\% for
the first stage, 7\% for the reduction stage, 62\% for the coincidence
stage, and 8\% in the output stage.

\section{Trigger Control System}
In our setup the function of the Trigger Control System (TCS) is to
distribute from a single source the trigger and time reference to the
readout modules. The TCS is based on the {\sc Compass} system, using a
laser module and passive optical splitters, which in turn is based on
the encoding method and distribution principles of the Time and
Trigger Control (TTC) system developed for the LHC
experiments~\cite{Abbon2007}.  A central component of the trigger
distribution is the TCS controller, a 6U VME module, that itself is
triggered by a first level trigger (FLT) from the {\sc Vuprom} system.
Our system consists of one TTC laser crate that operates at 155.52\,MHz,
two layers of optical 32-channel splitters and a total of 37 TCS
receivers. The receivers decode the trigger and deliver it to the
readout modules housed in the same VME crate.

\section{Data Acquisition System}
The architecture of the subsequent data acquisition (DAQ) system and
the synchronization of the read-out units are following the setup of
the {\sc Compass} experiment~\cite{Abbon2007}.

Upon arrival of the trigger signal the signal times are picked off by
144 TDC mezzanine boards that are plugged into the common read-out
driver modules developed for the {\sc Compass} experiment, named {\sc
  Catch}. These 9U VME modules are equipped with 4 slots for mezzanine
cards with 32 channels each. Each TDC mezzanine card hosts 4 dead-time
free ${\cal F}$1 chips, developed by the Faculty of Physics at the
University of Freiburg~\cite{COMPASS2001}. The ${\cal F}$1 TDC
performs digitization and readout asynchronously and without any dead
time.  The heart of the TDC is an asymmetric ring oscillator
representing the actual time in a bit value. The phased-locked-loop in
each of the ${\cal F}$1 is precisely synchronized to the 38.88\,MHz
reference clock provided by the TCS system.  The chips accept
falling, rising or both edges of the input signals, and stores them
with the time-stamps in an internal memory. The measuring unit has a
dynamic range of 62\,054 time steps. The length of an individual step
is determined by the clock frequency, $f$, the number of delay units
in the ring oscillator, $i=$ 19, and the settings of two pre-scalers,
$M=$ 23 and $N=$ 2~\cite{Braun1999}.  For the read-out of the fiber
detector, the normal resolution mode with 8 channels of $\Delta t= N
/ (f \cdot i \cdot M) \approx$ 118\,ps resolution (least significant
bit) has been chosen. The resolution is identical in all channels,
each channel has multi-hit capability, and the double pulse resolution
is typically 20\,ns.  With this setting a dynamic range of 62\,054
$\times$ 118\,ps $=$ 7.3\,$\mu$s can be used for the trigger signal
generation.

An incoming trigger signal is digitized by the same time measurement
unit as the hit-signals. A programmable trigger latency time is
subtracted from the measured trigger time to account for the time
needed to form and to distribute the trigger. A hit is considered to
coincide with a trigger if its time stamp is within a defined trigger
window following the latency corrected trigger time~\cite{Braun1999}.

The 37 {\sc Catch} modules combine the data of 32 $\times$ 4 $=$ 128
read-out channels each and transmit these sub-events via optical
S-LINK to read-out buffer PC (ROB) located in a radiation shielded
bunker. S-LINK multiplexer modules reduce the number of optical links.
The fiber detector read-out system comprises three 9U VME crates. Each
crate can be filled with a maximum of 16 {\sc Catch} modules providing
the readout of 128 $\times$ 12 $=$ 1\,536 channels.  The setup is
located $\sim$ 5\,m from the detector front-end, shielded from the
electromagnetic radiation by the yoke of the magnet and a shielding
wall.

The data acquisition, analysis and experiment control software used by
the A1 Collaboration consists of three major packages, {\sc Aqua},
{\sc Cola} and {\sc Mezzo}, written in {\sc C++} and
Java~\cite{Distler2001}. {\sc Mezzo} is used to operate and
monitor experimental devices, {\sc Cola} simulates and analyzes the
data, and the data acquisition is done by {\sc Aqua} (Data Acquisition for
A1 Experiments).  The {\sc Aqua} system is controlled through a GUI
(Graphics User Interface) on a host machine.  Before the start of a
run its conditions can be set in this interface. At the beginning of
each run, the program proceeds to set and read the slow control
diagnostics from {\sc Mezzo}. Also an automatic logbook entry is made
in the SQL database.

The ROB processors receive the data, pack it into {\sc Aqua}-defined
structures, and send it via a TCP/IP protocol to a central machine in
the counting room that collects data streams from different
spectrometer arms and stores it in run-files.  On this single machine,
data which belong to one physics event are combined in an event
building process and consistency checks are performed on the data
integrity. The event identification is achieved by matching a 16-bit
event number which includes a 3-bit event type definition.  When using
the spectrometer's electron-arm in coincidence with the hadron arm or
with one of the other spectrometers, the {\sc Catch} based DAQ is fully
integrated into the existing data acquisition without effecting the total
dead-time.

\begin{figure}
  \centering
  \includegraphics[width=0.94\columnwidth]{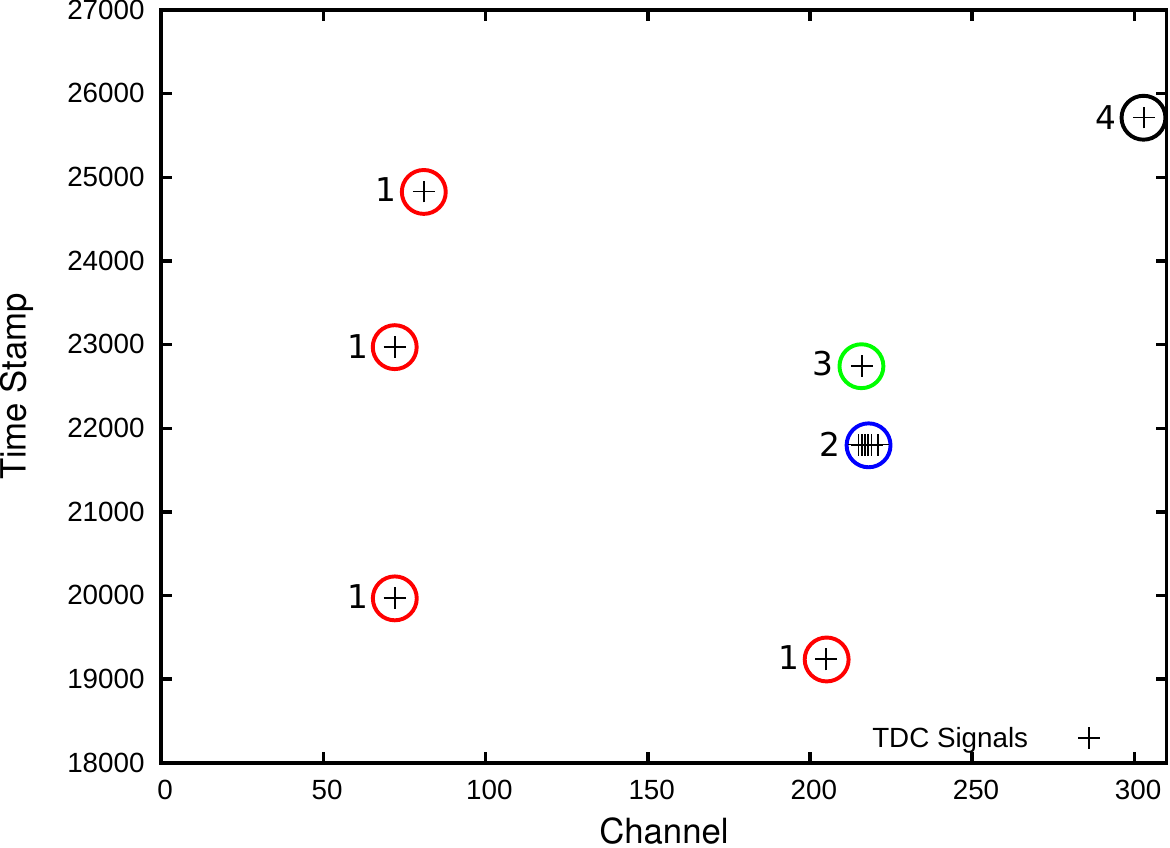}
  \includegraphics[width=\columnwidth]{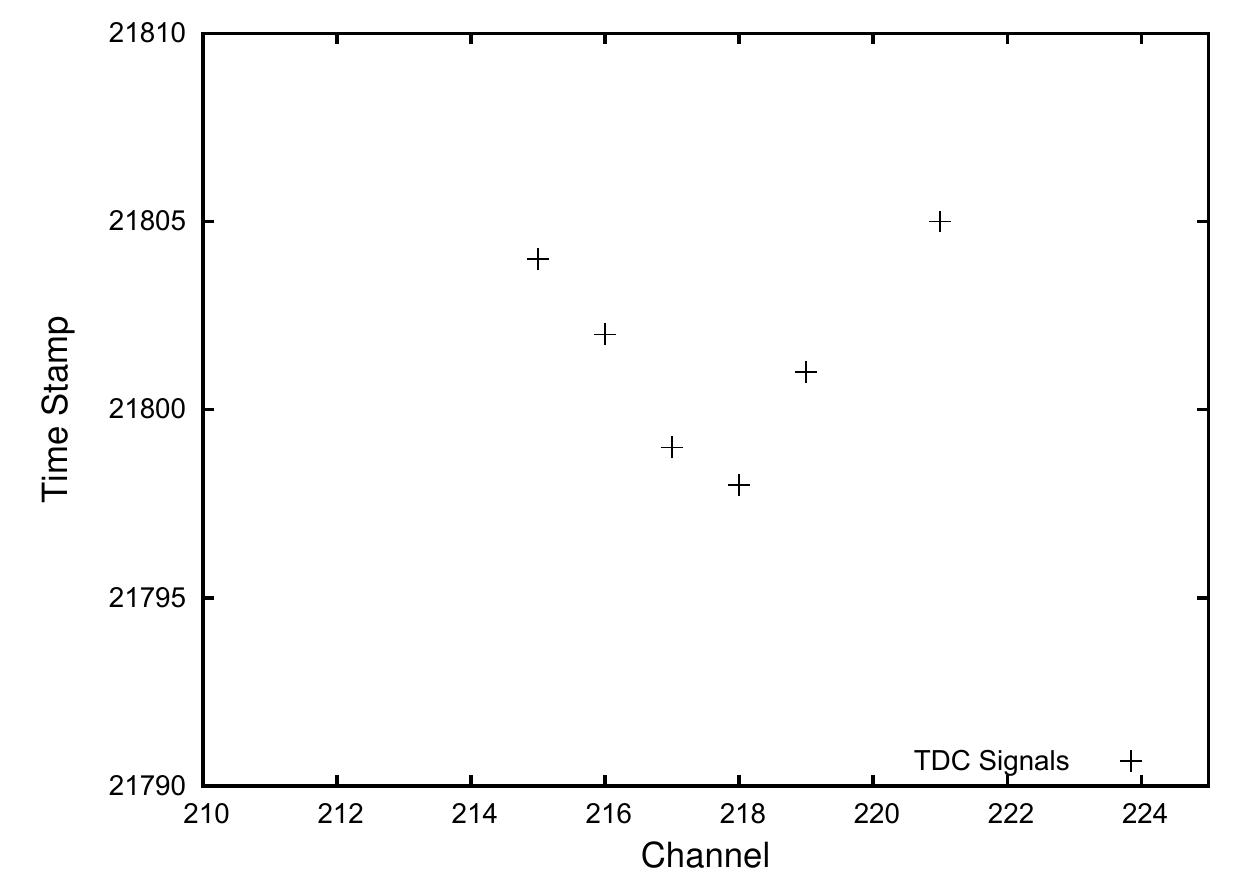}
  \caption{A typical event display of the fiber detector taken during
    the in-beam tests.  The data of 288 TDC channels is shown with
    each cross representing the leading edge time (in time bins of
    $\Delta t \approx$ 118\,ps) of a signal coming from the
    corresponding DTD channel. Different signal types are seen in this
    event: (1) single random hits, (2) a signal cluster with (3)
    after-pulse, and (4) the trigger reference time. A magnification
    of the signal cluster produced by the charged particle crossing is
    shown in the bottom panel. Further information on the cluster analysis
    is given in \cite{EsserDipl}.}
  \label{fig:events}
\end{figure}
%

\section{In-Beam Tests at MAMI}
\begin{figure*}
  \centering
  \includegraphics[width=0.8\textwidth]{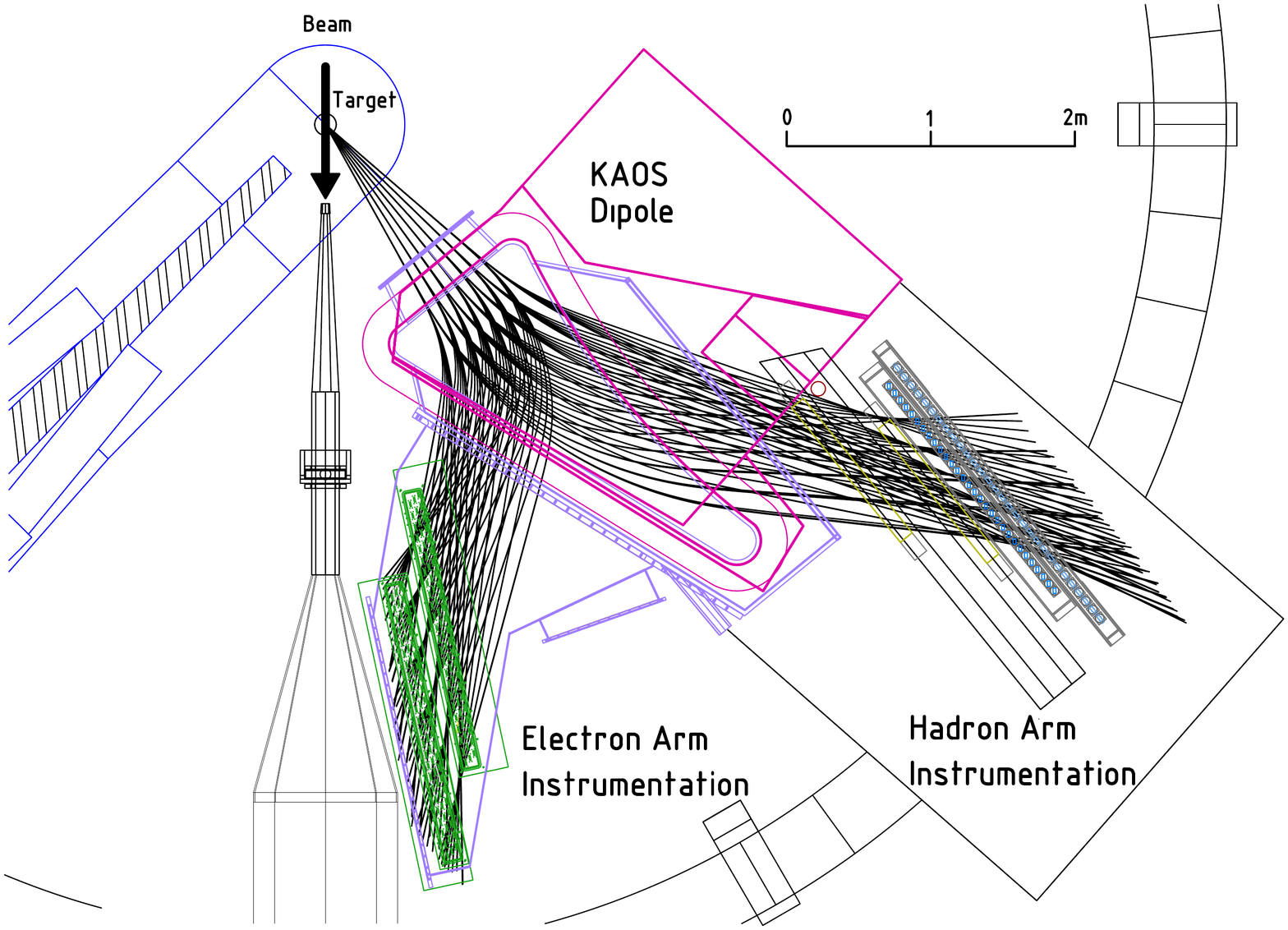}
  \caption{The \kaos\ double-arm spectrometer setup as used for the
    in-beam testing of its read-out and trigger electronics during
    2009. Ray-traced negatively and positively charged particle
    trajectories through the spectrometer are shown by full lines from
    target to detection planes. The position of the electron-arm
    detector and front-end boards is indicated in green color, its
    angle to the target is approx.~8--10$^\circ$.  In this setup the
    fringe field from the dipole coil deflects low momentum particles
    directly scattered into the direction of the electron-arm
    detector.}
  \label{fig:kaos-optics}
\end{figure*}

The read-out and trigger electronics for the electron-arm was
successfully tested during two beam-times in July and August 2009.
The \kaos\ spectrometer as realized in the spectrometer hall is
depicted schematically in Fig.~\ref{fig:kaos-optics}, where ray-traced
positively and negatively charged tracks for different momenta and
emission angles are shown.  The solid angle from the target covered by
the fiber detector was 6.4\,msr. A 585\,MeV electron beam was used on
a CH$_2$ target of 1.5\,mm (130\,mg$/$cm$^2$) thickness, resulting in
an electron--electron luminosity of ${\cal L} =$ 3.1 $\times$
10$^{34}$\,cm$^{-2}$s$^{-1}$ at a beam-current of 100\,nA.  Large
particle fluxes, dominated by M{\o}ller scattering of the electrons,
lead to high background rates in the fiber detectors at their exposed
position at 8--10$^\circ$. The contribution of particles inside the
acceptance of the spectrometer to the signal rate in the detector is
several orders of magnitude lower. Raw signal rates of more than 1\,MHz
were observed for beam-currents above 100\,nA. 
The measured rates for different trigger types are shown in 
Table~\ref{tab:triggerrates} for beam-currents of 10 and 100\,nA. 
Note that the beam-current measurement became imprecise for low currents.

\begin{table}[ht]
  \caption{Measured Trigger Rates in the Electron-Arm of the
    \kaos\ Spectrometer during Beam-Tests at MAMI in 2009. The
    Current of the 585\,MeV Electron Beam on a 130\,mg$/$cm$^2$ thick CH$_2$
    Target varied between 10 and 100\,nA.}
  \begin{center}
		\renewcommand{\arraystretch}{1.3}
    \begin{tabular}{lrr}
      \hline
      Trigger Type & Trigger Rate (kHz) & Trigger Rate (kHz)\\
      				     & @ 10\,nA & @ 100\,nA\\
      \hline
      Raw Signal Rate           & 100 $\pm$ 100 & 1\,120  $\pm$ 30   \\
      Clusters in $x$           & 7.33 $\pm$ 0.09 & 47.0  $\pm$ 0.2  \\
      Clusters in $\theta$      & 5.91 $\pm$ 0.08 & 37.3  $\pm$ 0.2  \\
      $x$ OR $\theta$           & 13.2 $\pm$ 0.1  & 83.9  $\pm$ 0.3  \\
      $x$ AND $\theta$          & (63 $\pm$ 8) $\times$ 10$^{-3}$  & (490 $\pm$ 20) $\times$ 10$^{-3}$ \\
      Random Coincidences       & (3.46 $\pm$ 0.06) $\times$ 10$^{-3}$ & (140 $\pm$ 1) $\times$ 10$^{-3}$\\
      \hline
    \end{tabular}
  \end{center}
  \label{tab:triggerrates}
\end{table}
\begin{figure}
  \centering
  \includegraphics[angle=90,width=\columnwidth]{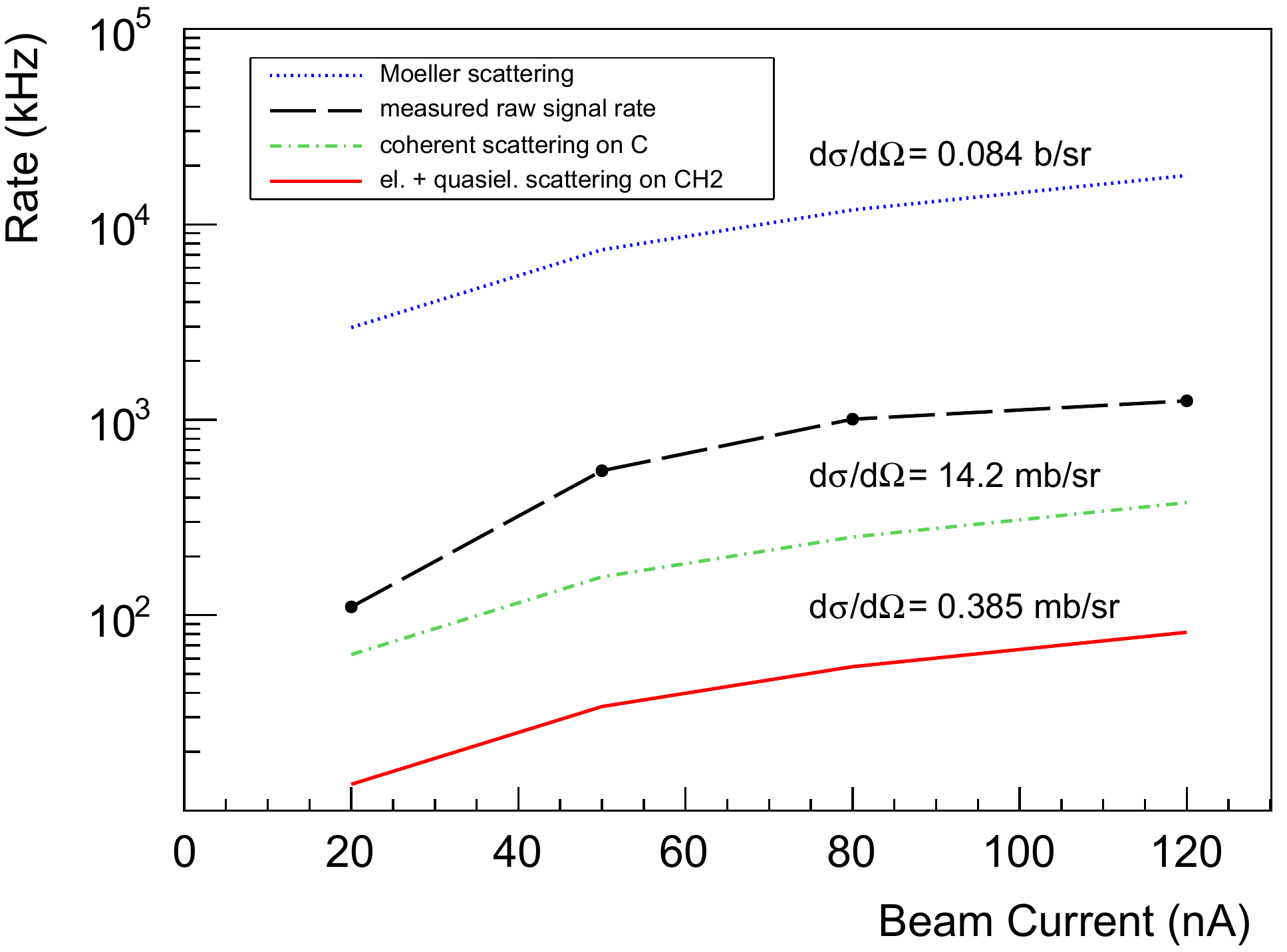}
  \caption{Measured raw signal rates in the $x$-plane of the electron-arm
    instrumentation as a function of electron beam current compared to the
    calculated scattering rates of (M{\o}ller scattering,
    elastic$/$quasi-elastic scattering at H and C nuclei, and coherent
    scattering at C nuclei). Values for the differential cross-sections are 
    shown as well.}
  \label{fig:rates}
\end{figure}

Fig.~\ref{fig:events} shows a typical event of the fiber detector
taken during the in-beam tests. The data of 288 TDC channels is shown
with each cross representing the leading edge time (in time steps of
$\Delta t \approx$ 118\,ps) of a signal coming from the corresponding
DTD channel. Signal types are classified into (1) single random hits,
(2) signal clusters, (3) after-pulses, and (4) trigger reference
time. A magnification of the signal cluster is shown in the bottom
panel. The cluster is spread over 7 channels and 7 time steps,
corresponding to 0.8\,ns. It is known that charged particles crossing
the fiber detector produce the latter type of
signals~\cite{Achenbach2008:SciFi}. In the analysis software the
clusters were identified based on a hierarchical agglomerative cluster
algorithm. In deducing the hit times an iteration over all hits in a
plane including multiple hits in a channel was performed, and clusters
of correlated hit times were searched for. The cluster with the time
closest to the trigger signal time was retained, and within the
cluster the time of the leading signal was chosen as hit time. A
minimum temporal and spatial distance between signals in a cluster
have been defined before the analysis took place.

In Fig.~\ref{fig:rates} the raw signal rate in the $x$-plane of the
electron-arm detector is compared to the calculated rates of M{\o}ller
scattering, elastic$/$quasi-elastic scattering at H and C nuclei, and coherent
scattering at C nuclei.  The momentum of M{\o}ller electrons scattered into
the direction of the electron-arm detector was only $p\approx$
40--50\,MeV$\!/c$, so most of these electrons were deflected in the
fringe field of the dipole, {\em cf.}  Fig.~\ref{fig:kaos-optics}, or
scattered on the vacuum chamber walls.  Taking these losses into
account, the measured rate is in good agreement with the calculated
rates for scattering on the nuclei with a contribution from M{\o}ller
scattering.  Signals of type (1) or (3) could be filtered out by the
clustering algorithm of the trigger system. The rate then dropped by a
factor 10. The tests have shown that a reliable triggering on signal
clusters is achieved with the {\sc Vuprom} boards.

The background reduction in this spectrometer geometry can be deduced from 
the numbers shown in Table~\ref{tab:triggerrates} and can be factored
into two parts. The first factor was determined
to be 1~:~10 requiring the right cluster sizes and therefore rejecting
direct scattering products. The second factor was measured to be
1~:~200 requiring coincident events in both planes and therefore rejecting
a high degree of particle background. The trigger logic
is inefficient only for events with simultaneous temporal and spatial pile-up, 
{\em i.e.} when a background particle hits the same cluster at the same time period
as a particle coming through the spectrometer. For the in-beam tests such events
could be neglected. In a later extension, when the logic will include conditions on the 
reaction kinematics,
a trigger efficiency $<$ 100\% will be inevitable. To avoid the generation
of fake triggers in the first stage all signals are passed through 
pulse-width discriminators so that the cluster size determination is not
affected by asynchronous signals.

The performance of the full trigger system is currently under study 
using further test-beams at MAMI and dedicated setups in the laboratory.

\section{Conclusion}
An FPGA-based trigger and DAQ system with more than 4\,000 readout
channels has been installed for the fast signals of the fiber detector
in the \kaos\ spectrometer's electron-arm. The system was operating
successfully in several beam-times where sophisticated trigger
conditions have distinguished background particles from 
particles passing through the spectrometer. The reduction of the raw
signal rate to the trigger rate was of the order 1~:~2\,000 for the
conditions experienced in the tests.

\section*{Acknowledgment}
We deeply appreciate the support of our colleagues from the {\sc Compass}
collaboration and the staffs of the collaborating institutions
involved in the DAQ electronics development.

\IEEEtriggeratref{7}

\end{document}